\documentclass{acmtrans2e}

\usepackage{rotating}
\usepackage{graphicx}

\newtheorem{theorem}{Theorem}

\title{Optimal Lower and Upper Bounds \\ for Representing Sequences}

\author{%
DJAMAL BELAZZOUGUI \\
Helsinki Institute for Information Technology HIIT and
University of Helsinki \\
GONZALO NAVARRO \\
University of Chile}

\begin{abstract}
Sequence representations supporting queries $access$, $select$ and $rank$
are at the core of many data structures. There is a considerable gap between
the various upper bounds and the few lower bounds known for such  
representations, and how they relate to the space used. In this article
we prove a strong lower bound for $rank$, which holds for rather
permissive assumptions on the space used, and give matching upper bounds that 
require only a compressed representation of the sequence. Within this
compressed space, operations $access$ and $select$ can be solved in
constant or almost-constant time, which is optimal for large alphabets.
Our new upper bounds dominate all of the previous work in the 
time/space map.
\end{abstract}



\begin{document}

\maketitle

\begin{bottomstuff}
Partially funded by Fondecyt Grant 1-110066, Chile.
First author also partially supported by 
the French ANR-2010-COSI-004 MAPPI Project and
the Academy of Finland under grant 250345 (CoECGR).

An early partial version of this work appeared in 
{\em Proc. ESA'12} \cite{BN12}.

Authors' address: Djamal Belazzougui,
Helsinki Institute for Information Technology HIIT, 
Department of Computer Science, University of Helsinki, 
Finland,
{\tt djamal.belazzougui@gmail.com}.
Gonzalo Navarro,
Department of Computer Science, University of Chile, Chile,
\texttt{gnavarro@dcc.uchile.cl}
\end{bottomstuff}

\section{Introduction}

A large number of data structures build on sequence representations. In 
particular, supporting the following three queries on a sequence $S[1,n]$
over alphabet $[1,\sigma]$ has proved extremely useful:
\begin{itemize}
\item $access(S,i)$ gives $S[i]$;
\item $select_a(S,j)$ gives the position of the $j$th occurrence of 
$a \in [1,\sigma]$ in $S$; and
\item $rank_a(S,i)$ gives the number of occurrences of $a \in [1,\sigma]$
in $S[1,i]$.
\end{itemize}

The most basic case is that of bitmaps, when $\sigma=2$. Obvious applications 
are set representations supporting membership and predecessor search, although
many other uses, such as representing tree topologies, multisets, and partial
sums \cite{Jac89,RRR07} have been reported. The focus of this article is on
general alphabets, where further applications have been described.
For example, the FM-index \cite{FM05}, a compressed indexed
representation for text collections that supports pattern searches, is most
successfully implemented over a sequence representation supporting $access$ 
and $rank$ \cite{FMMN07}, and more recently $select$ \cite{BN11}. 
\citeN{GGV03} had used earlier similar techniques for text 
indexing. \citeN{GMR06} used these operations for representing 
labeled trees and permutations. Further applications of these operations to 
multi-labeled trees and binary relations were uncovered by 
\citeN{BHMR11}. \citeN{FLMM09}, \citeN{GHSV06},
and \citeN{ACMNNSV10}
 devised new applications to XML indexing. 
Other applications were described as well to representing permutations and
inverted indexes \cite{BN09,BCGNN12} and graphs \cite{CN10,HN12}.
\citeN{VM07} and \citeN{GNP10} applied them to document retrieval on general 
texts. Finally, applications to various types of inverted indexes on natural 
language text collections have been explored \cite{BFLN12,AGO10,AGMOS12}.

When representing sequences supporting the three operations, it seems reasonable
to aim for $O(n\lg\sigma)$ bits of space. However, in many applications the
size of the data is huge and space usage is crucial: only sublinear space on 
top of the raw data can be accepted. This is our focus.

Various time- and space-efficient sequence representations supporting the 
three operations have been proposed, and also various lower bounds have been
proved. All the representations proposed assume the RAM model with 
word size $w=\Omega(\lg n)$. 
In the case of bitmaps, \citeN{Mun96} and \citeN{Cla96} achieved constant-time 
$rank$  and $select$ using $o(n)$ extra bits on top of a plain representation 
of $S$. \citeN{Gol07} proved a lower bound of $\Omega(n\lg\lg n/\lg n)$ extra
bits for supporting either operation in constant time if $S$ is to be 
represented in plain form, and gave matching upper bounds. This assumption
is particularly
inconvenient in the frequent case where the bitmap is sparse, that is, it has 
only $m \ll n$ 1s, and hence can be compressed. When $S$ can be
represented arbitrarily, \citeN{Pat08} achieved $\lg{n\choose m}
+ O(n/\lg^c n)$ bits of space, where 
$c$ is any constant. This space was shown later to be optimal \cite{PV10}. 
However, the space can be reduced further, up to $\lg {n \choose m} + O(m)$ 
bits, if superconstant time for the operations is permitted \cite{GHSV07,OS07},
or if the operations are weakened: When $rank_1(S,i)$ can only be applied if 
$S[i]=1$ and only $select_1(S,j)$ is supported, \citeN{RRR07} 
achieved constant time and $\lg {n \choose m} + o(m) + O(\lg\lg n)$ bits of 
space. When only $rank_1(S,i)$ is supported for the positions $i$ such that
$S[i]=1$, and in addition we cannot
even determine $S[i]$, the structure is called a monotone minimum perfect
hash function (mmphf) and can be implemented in $O(m\lg\lg\frac{n}{m})$ bits
and answering in constant time \cite{BBPV09}.

For general sequences, a
useful measure of compressibility is the {\em zeroth-order entropy} of
$S$, $H_0(S) = \sum_{a \in [1,\sigma]} \frac{n_a}{n}\lg\frac{n}{n_a}$, where 
$n_a$ is the number of occurrences of $a$ in $S$. This can be extended to the
{\em $k$-th order entropy}, $H_k(S) = \frac{1}{n} \sum_{A \in [1,\sigma]^k}
|T_A| H_0(T_A)$, where $T_A$ is the string of symbols following $k$-tuple $A$
in $S$. It holds $0 \le H_k(S) \le H_{k-1}(S) \le H_0(S) \le \lg\sigma$ for
any $k$, but the entropy measure is only meaningful for $k < \lg_\sigma n$.
See \citeN{Man01} and \citeN{Gag06} for a deeper discussion.

We say that a representation of $S$ is {\em succinct} if it takes $n\lg\sigma +
o(n\lg\sigma)$ bits, {\em zeroth-order} compressed if it takes 
$nH_0(S)+o(n\lg\sigma)$ bits, and {\em high-order} compressed 
if it takes $nH_k(S)+o(n\lg\sigma)$ bits. We may also 
compress the redundancy, $o(n\lg\sigma)$, to use for example 
$nH_0(S)+o(nH_0(S))$ bits.

Upper and lower bounds for sequence representations supporting the three
operations are far less understood over arbitrary alphabets. 
\citeN{GGV03} introduced the {\em wavelet tree}, a zeroth-order compressed 
representation
using $nH_0(S)+o(n\lg\sigma)$ bits that solves the three queries in time 
$O(\lg\sigma)$. The time was reduced to $O\left(1+\frac{\lg\sigma}{\lg\lg n}\right)$ with 
multiary wavelet trees \cite{FMMN07}, and later the space was reduced to
$nH_0(S)+o(n)$ bits \cite{GRR08}. Note that the query times are constant for
$\lg\sigma = O(\lg\lg n)$, that is, $\sigma = O(\mathrm{polylog}~n)$.
\citeN{GMR06} proposed a succinct representation that is more
interesting for large alphabets. It solves
$access$ and $select$ in $O(1)$ and $O(\lg\lg\sigma)$ time, or vice versa,
and $rank$ in $O(\lg\lg\sigma)$ time or slightly more. This representation 
was made slightly faster (i.e., $rank$ time is always $O(\lg\lg\sigma)$)
and compressed to $nH_0(S)+o(nH_0(S))+o(n)$ by \citeN{BCGNN12}.
Alternatively, \citeN{BHMR11} achieved high-order compression,
$nH_k(S)+o(n\lg\sigma)$ bits for any $k=o(\lg_\sigma n)$, and slightly higher 
times, which were again reduced by \citeN{GOR10}.

There are several curious aspects in the map of the current solutions for 
general sequences. On the one hand, in various solutions for large alphabets
\cite{GMR06,BCGNN12,GOR10} the times for $access$ and $select$ seem 
to be complementary (i.e., one is constant and the other is not), whereas that 
for $rank$ is always superconstant. On the other hand, there is no smooth 
transition between the complexity of the wavelet-tree based solutions, 
$O\left(1+\frac{\lg\sigma}{\lg\lg n}\right)$, and those for larger alphabets, 
$O(\lg\lg\sigma)$.

The complementary nature of $access$ and $select$ is not a surprise. 
\citeN{Gol09} proved lower bounds that relate the time performance that can be 
achieved for these operations with the redundancy of any encoding 
of $S$ on top of its information content. The lower bound acts on the product 
of both times, that is, if $t$ and $t'$ are the time complexities for $access$
and $select$, and $\rho$ 
is the bit-redundancy per symbol, then $\rho \cdot t \cdot t' = 
\Omega((\lg\sigma)^2/w)$ holds for a wide range of values of $\sigma$.
Many upper bounds for large alphabets \cite{GMR06,BCGNN12,GOR10} match this 
lower bound when $\lg\sigma=\Theta(w)$.

Despite operation $rank$ seems to be harder than the others (at least no
constant-time solution exists except for polylog-sized alphabets), no general
lower bounds on this operation have been proved. Only a result 
\cite{GOR10} for the
case in which $S$ must be encoded in plain form states that if one solves
$rank$ within $a = O\left(1+\frac{\lg\sigma}{\lg\lg\sigma}\right)$ accesses to the sequence,
then the redundancy per symbol is $\rho = \Omega((\lg\sigma)/a)$.
Since in the RAM model one can access up to $w/\lg\sigma$ symbols in one
access, this implies a lower bound of $\rho \cdot t = \Omega((\lg\sigma)^2/w)$,
similar to the one by \citeN{Gol09} for the product of $access$ and
$select$ times and also matched by current solutions \cite{GMR06,BCGNN12,GOR10}
when $\lg\sigma=\Theta(w)$.

In this article we make several contributions that help close the gap between
lower and upper bounds on sequence representation.

\begin{enumerate}
\item We prove the first general lower bound on $rank$, which shows that this
operation is, in a sense, noticeably harder than the others: Any structure 
using $O(n \cdot w^{O(1)})$ bits needs time 
$\Omega\left(\lg\frac{\lg\sigma}{\lg w}\right)$ to answer $rank$ queries 
(the bound is only $\Omega\left(\lg\frac{\lg n}{\lg w}\right)$ if $\sigma>n$;
we mostly focus on the interesting case $\sigma \le n$). 
Note that the space includes the rather permissive
$O(n\cdot \mathrm{polylog}~n)$. The existing lower bound \cite{GOR10} not only 
is restricted to plain encodings of $S$ but only forbids achieving this time
complexity within $n\lg\sigma + O\left(n\lg^2\sigma/(w\lg\frac{\lg\sigma}{\lg w})\right)
= n\lg\sigma + o(n\lg\sigma)$ bits of space.
Our lower bound uses a reduction from predecessor queries \cite{PT08}.
\item We give a matching upper bound for $rank$, using $O(n\lg\sigma)$ bits
of space and answering queries in time $O\left(\lg\frac{\lg\sigma}{\lg w}\right)$.
This is lower than any time complexity achieved so far for this operation
within $O(n \cdot w^{O(1)})$ bits, and it elegantly unifies both 
known upper bounds under a single and lower time complexity. This is achieved 
via a reduction to a predecessor query structure that is tuned to use slightly 
less space than usual.
\item We derive succinct and compressed representations of sequences
that achieve time $O\left(1+\frac{\lg\sigma}{\lg w}\right)$ for $access$, $select$ and 
$rank$, improving upon previous results \cite{FMMN07,GRR08}. This yields
constant-time operations for $\sigma = w^{O(1)}$. Succinctness is achieved by 
replacing universal tables used in previous solutions \cite{FMMN07,GRR08} with 
bit manipulations in the RAM model. Compression is achieved by combining the 
succinct representation with known compression boosters \cite{BCGNN12}.
\item We derive succinct and compressed representations 
of sequences over larger alphabets, which achieve the optimal time 
$O\left(\lg\frac{\lg\sigma}{\lg w}\right)$ for $rank$, and 
almost-constant time for $access$ and $select$ (i.e., one is constant time and
the other any superconstant time, as low as desired). The result improves upon 
all succinct and compressed representations proposed so far
\cite{GMR06,BHMR11,BCGNN12,GOR10}. This is achieved by plugging our 
$O(n\lg\sigma)$-bit solutions into some of those succinct and compressed data 
structures.
\item As an immediate application, we obtain the fastest text self-index
\cite{GGV03,FM05,FMMN07} able to provide pattern matching on a text compressed 
to its $k$th order entropy within $o(n)(H_k(S)+1)$ bits of redundancy, improving
upon the best current one \cite{BCGNN12}, and being only slightly slower than 
the fastest one \cite{BN11}, which however poses $O(n)$ further bits of space
redundancy.
\end{enumerate}

Table~\ref{tab:previous} compares our new upper bounds with the best current 
ones. It can be seen that, combining our results, we dominate all of the
best current work \cite{GRR08,BCGNN12,GOR10}, as well as earlier ones
\cite{GMR06,FMMN07,BHMR11} (but our solutions build on some of those). 

\begin{sidewaystable}[tp]
  \caption{
The best previous upper bounds, and our new best ones, for data structures 
supporting $access$, $select$ and $rank$.
The space bound $H_k(S)$ holds for any \(k = o (\lg_\sigma n)\).}
\label{tab:previous}
\medskip
\small
\begin{center}
{\begin{tabular}
{c@{\hspace{1ex}}|@{\hspace{2ex}}c@{\hspace{3ex}}c@{\hspace{3ex}}c@{\hspace{3ex}}c@{\hspace{1ex}}c}
source & space (bits)                                                & $access$                              & $select$                                        & $rank$                                  \\[1ex]
\hline
& & & \\[-1ex]
\cite[Thm.~4]{GRR08}   & \(n H_0(S) + o(n)\)                              & $O\left(1+\frac{\lg \sigma}{\lg \lg n}\right)$ & $O\left(1+\frac{\lg \sigma}{\lg \lg n}\right)$        & $O\left(1+\frac{\lg \sigma}{\lg \lg n}\right)$    \\[1ex]
\cite[Thm.~2]{BCGNN12} & \(n H_0(S) + o(nH_0(S))+o(n)\)                               & $O(\lg \lg \sigma)$                 & $O(1)$
	    & $O(\lg \lg \sigma)$ \\[1ex]
\cite[Thm.~2]{BCGNN12} & \(n H_0(S) + o(nH_0(S))+o(n)\)                               & $O(1)$                              & $O(\lg \lg \sigma)$ & $O(\lg \lg \sigma)$                    \\[1ex]
\cite[Cor.~2]{GOR10}   & \(n H_k(S) + o(n\lg \sigma)\) & $O(1)$ & $O(\lg\lg\sigma)$ & $O(\lg\lg\sigma)$ \\[1ex] %OJO can be any omega_sigma(1) \\[1ex]
\hline 
& & & \\[-1ex]
Theorem~\ref{thm:upper-smallH0} & $nH_0(S)+o(n)$ 
                 & $O\left(1+\frac{\lg \sigma}{\lg w}\right)$ & $O\left(1+\frac{\lg \sigma}{\lg w}\right)$        & $O\left(1+\frac{\lg \sigma}{\lg w}\right)$    \\[1ex]
Theorem~\ref{thm:upper-H0} & $nH_0(S)+o(nH_0(S))+o(n)$
		& any $\omega(1)$ & $O(1)$ & $O\left(\lg\frac{\lg\sigma}{\lg w}\right)$ \\[1ex]
Theorem~\ref{thm:upper-H0} & $nH_0(S)+o(nH_0(S))+o(n)$
		& $O(1)$ & any $\omega(1)$ & $O\left(\lg\frac{\lg\sigma}{\lg w}\right)$ \\[1ex]
Theorem~\ref{thm:upper-Hk} ($\lg\sigma=\omega(\lg w)$) & $nH_k(S) + o(n\lg\sigma)$
		& $O(1)$ & any $\omega(1)$ & $O\left(\lg\frac{\lg\sigma}{\lg w}\right)$ \\[1ex]
Theorem~\ref{thm:lower-Hk} ($\lg\sigma=O(\lg w)$) & $nH_k(S) + o(n\lg\sigma)$
		& $O(1)$ & any $\omega(1)$ & any $\omega(1)$ \\[1ex]
\end{tabular}}
\end{center}
\end{sidewaystable}

Besides $w=\Omega(\lg n)$, we make for simplicity the reasonable assumption 
that $\lg w = O(\lg n)$, that is, $w = n^{O(1)}$; this avoids irrelevant 
technical issues (otherwise, for example, all the text fits in a single 
machine word!). We also avoid mentioning the need to store a constant number
of systemwide pointers ($O(w)$ bits), which is needed in any reasonable
implementation. Finally, our results assume that, in the RAM model, bit shifts, 
bitwise logical 
operations, and arithmetic operations (including multiplication) are permitted. 
Otherwise we can simulate them with universal tables using $o(2^w)$ extra bits
of space. This space is $o(n)$ if $\lg n \ge w+O(1)$; otherwise we can reduce 
the universal tables to use $o(n)$ bits, but any $\lg w$ in the upper bounds 
becomes $\lg\lg n$.

The next section proves our lower bound for $rank$. 
Section~\ref{sec:upper} gives a matching upper bound within $O(n\lg\sigma)$
bits of space. Within this space, achieving constant time for $access$
and $select$ is trivial.
Section~\ref{sec:succinct} shows how to retain the same upper bound for
$rank$ within succinct space, while reaching constant or almost-constant time 
for $access$ and $select$.
Section~\ref{sec:compressed} retains those times while reducing the size
of the representation to zeroth-order or high-order compressed space.
Finally, Section~\ref{sec:concl} gives our conclusions and future challenges.

\section{Lower Bound for Rank}

Our technique is to reduce from a predecessor problem and apply the 
density-aware lower bounds of \citeN{PT06}.
Assume that we have $n$ keys from a universe of size $u = n\sigma$, then the 
keys are of length $\ell = \lg u = \lg n + \lg \sigma$. According to branch 2
of \citeANP{PT06}'s result, the time for predecessor queries in this 
setting is lower bounded
by $\Omega\left(\lg \left(\frac{\ell - \lg n}{a}\right)\right)$, where
$a = \lg (s/n)+\lg w$ and $s$ is the space in words of our representation 
(the lower bound is in the cell probe model for word length $w$, so 
the space is always expressed in number of cells). The lower bound holds even 
for a more restricted version of the predecessor problem in which one of two 
colors is associated with each element and the query only needs to return the 
color of the predecessor. 

The reduction is as follows. 
We divide the universe $[1,n\cdot \sigma]$ into 
$\sigma$ intervals, each of size $n$. This division can be viewed as a binary
matrix of $n$ columns $c \in [1,n]$ and $\sigma$ rows $r \in [1,\sigma]$, 
where we set a 1 at row $r$ and 
column $c$ iff element $(r-1)\cdot n+c$ belongs to the set.
We will use four data structures.

\begin{enumerate}
\item A plain bitvector $L[1,n]$ which stores the color associated 
with each element. The array is indexed by the original ranks of the elements.

\item
A partial sums structure $R$ stores the number of elements in each row. 
This is a bitmap concatenating the $\sigma$ unary representations,
$1^{n_r}0$, of the number $n_r$ of 1s in each row $r \in [1,\sigma]$. Thus $R$
is of length $n+\sigma$ and can give in constant time the number of 1s up
to (and including) any row $r$, 
$count(r) = rank_1(R,select_0(R,r)) = select_0(R,r)-r$, in constant
time and $O(n+\sigma)$ bits of space \cite{Mun96,Cla96}.

\item
A column mapping data structure $C$ that maps the original columns into a set
of columns where $(i)$ empty columns are eliminated, and $(ii)$ new columns
are created when two or more 1s fall in the same column. $C$ is a bitmap
concatenating the $n$ unary representations, $1^{n_c}0$, of the number $n_c$
of 1s in each column $c \in [1,n]$. So $C$ is of length $2n$. Note that the 
new matrix of mapped columns also has $n$ columns (one per element in the set) 
and exactly one 1 per column. The original column $c$ is then mapped to 
$col(c) = rank_1(C,select_0(C,c)) = select_0(C,c)-c$, using constant time
and $O(n)$ bits. Note that $col(c)$ is the last of the columns to which the
original column $c$ might have been expanded.

\item
A string $S[1,n]$ over alphabet $[1,\sigma]$, so that $S[c]=r$ iff the only
1 at column $c$ (after column remapping) is at row $r$. Over this string we 
build a data structure able to answer queries $rank_r(S,c)$.

\end{enumerate}

Colored predecessor
queries are solved in the following way. Given an element $x\in [1,u]$, we 
first decompose it into a pair $(r,c)$ where $x=(r-1)\cdot n+c$ and $1\le c\le n$.
In a first step, we compute $count(r-1)$ in constant time. This gives us the 
count of elements up to point $(r-1)\cdot n$. Next we must compute the count of 
elements in the range $[(r-1)\cdot n+1,(r-1)\cdot n+c]$. 
For doing that we first remap the column to $c' = col(c)$ in constant time, 
and finally compute $rank_r(S,c')$, which gives the number of 1s in row $r$
up to column $c'$. Note that if column $c$ was expanded to several ones, we
are counting the 1s up to the last of the expanded columns, so that all the
original 1s at column $c$ are counted at their respective rows.
Then the rank of the predecessor of $x$ is $p=count(r-1)+rank_r(S,col(c))$.
Finally, the color associated with $x$ is given by $L[p]$.
%We can then associate any information to it in an array indexed by such rank;
%in our case, the identity of the element (say, using $w$ bits).

\begin{figure}[t]
\begin{center}
\includegraphics[width=0.6\textwidth]{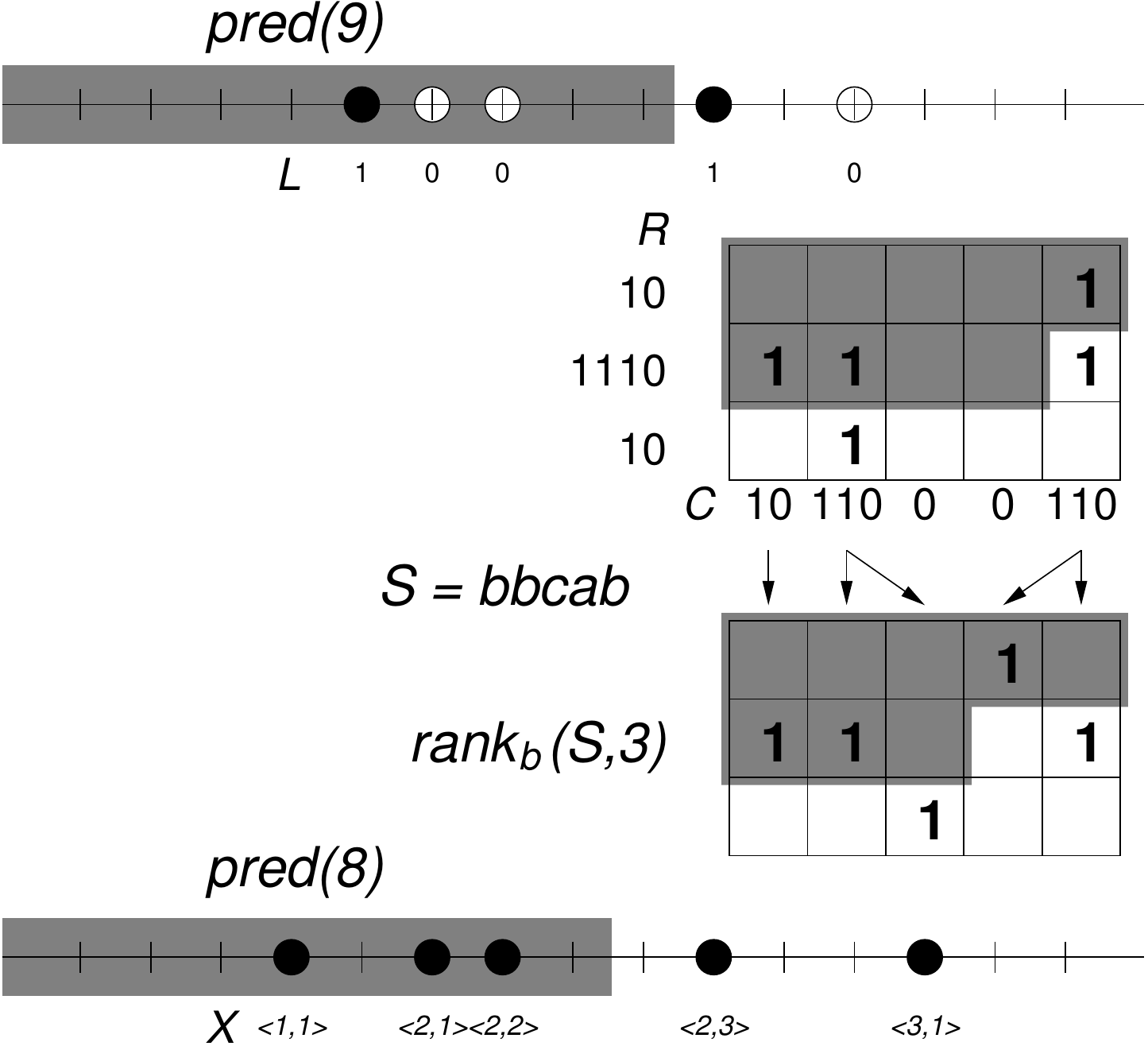}
\end{center}
\caption{Illustration of the lower bound technique, moving from a predecessor
to a $rank$ query, and of the upper bound technique, moving from the $rank$ 
query to a predecessor query.}
\label{fig:lbub}
\end{figure}

\paragraph*{Example} Fig.~\ref{fig:lbub} illustrates the technique on 
a universe of size $u=n \times \sigma = 5 \times 3 = 15$, the set 
$\{ 5,6,7,10,12\}$ of $n=5$ points black or white, and the query $pred(9)$, 
which must return the color of the 3rd point. The bitmap $L$ indicates the
colors of the points. The top matrix is obtained by taking the first, second,
and third ($\sigma=3$) segments of length $n=5$ from the universe, and 
identifying points with 1-bits (the omitted cells are 0-bits). Bitmaps $R$
and $C$ count the number of 1s in rows and columns, respectively. Bitmap $C$
is used to map the matrix into a new one below it, with exactly one point per
column. Then the predecessor query is mapped to the matrix, and spans several
whole rows (only 1 in this example) and one partial row. The 1s in whole rows 
(1 in total) are counted 
using $R$, whereas those in the partially covered row are counted with a 
$rank_b(S,3)=2$ query on the string $S=bbcab$ represented by the mapped matrix.
Then we obtain the desired 3 (3rd point), and $L[3]=0$ is the color.
Ignore for now the last line in the figure. \hfill $\Box$

\begin{theorem}
Given a data structure that supports $rank$ queries on strings of length
$n$ over alphabet $[1,\sigma]$, in time $t(n,\sigma)$ and using 
$s(n,\sigma)$ bits of space, we can solve the colored predecessor problem for 
$n$ integers from universe $[1,n\sigma]$ in time $t(n,\sigma)+O(1)$ using a 
data structure that occupies $s(n,\sigma)+O(n+\sigma)$ bits.
\end{theorem}

By the reduction above we get that any lower bound for predecessor search for 
$n$ keys over a universe of size $n\sigma$ must also apply to $rank$ queries 
on sequences of length $n$ over alphabet of size $\sigma$. In our case, if we 
aim at using $n \cdot w^{O(1)}$ bits of space for the $rank$ data structure, 
and allow any $\sigma \le n \cdot w^{O(1)}$, this lower bound (branch 2 
\cite{PT06}) is $\Omega\left(\lg\frac{\ell-\lg n}{\lg(s/n)+\lg w}\right) =
 \Omega\left(\lg\frac{\lg\sigma}{\lg w}\right)$.

\begin{theorem} \label{thm:lb}
Any data structure that uses $n\cdot w^{O(1)}$ space to represent 
a sequence of length $n$ over alphabet $[1,\sigma]$, for any 
$\sigma \le n \cdot w^{O(1)}$, must use time 
$\Omega\left(\lg\frac{\lg\sigma}{\lg w}\right)$ to answer $rank$ 
queries. 
\end{theorem}

For larger $\sigma$, the space of our representation is dominated by the
$O(\sigma)$ bits of structure $R$, so the lower bound becomes
$\Omega\left(\lg\frac{\lg\sigma}{\lg(\sigma/n)}\right)$, which 
worsens (decreases) as $\sigma$ grows from $n\cdot w^{\omega(1)}$, and
becomes completely useless for $\sigma = n^{1+\Omega(1)}$.
However, since the time for $rank$ is monotonic in $\sigma$, we still
have the lower bound $\Omega\left(\lg\frac{\lg n}{\lg w}\right)$
when $\sigma > n$; thus a general lower bound is
$\Omega\left(\lg\frac{\lg \min(\sigma,n)}{\lg w}\right)$ time.
For simplicity we have focused in the most interesting case.

Assume to simplify that $w = \Theta(\lg n)$.
The lower bound of Theorem~\ref{thm:lb}
is trivial for small $\lg\sigma=O(\lg\lg n)$ (i.e.,
$\sigma = O(\mathrm{polylog}~n)$), where constant-time solutions for 
{\em rank} exist that require only $nH_0(S)+o(n)$ bits \cite{FMMN07}. 
On the other hand, if $\sigma$ is sufficiently large, 
$\lg\sigma = (\lg\lg n)^{1+\Omega(1)}$,
the lower bound becomes simply $\Omega(\lg\lg\sigma)$, where it is matched
by known compressed solutions requiring as little as 
$nH_0(S)+o(nH_0(S))+o(n)$ \cite{BCGNN12} or $nH_k(S)+o(n\lg\sigma)$ 
\cite{GOR10} bits.

The range where this lower bound has not yet been matched is 
$\omega(\lg\lg n) = \lg\sigma = (\lg\lg n)^{1+o(1)}$.
It is also unmatched when $\lg n = o(w)$.
The next section presents a new matching upper bound.

\section{Optimal Upper Bound for Rank} \label{sec:upper}

We now show a matching upper bound with optimal time and space 
$O(n\lg\sigma)$ bits. In the next sections we make the space succinct and even
compressed.

We reduce the problem to predecessor search and then use a convenient solution
for that problem.
The idea is simply to represent the string $S[1,n]$ over alphabet $[1,\sigma]$
as a matrix of $\sigma$ rows and $n$ columns, and regard each $S[c]$ as a
point $(S[c],c)$. Then we represent the matrix as 
the set of $n$ points $\{ (S[c]-1) \cdot n + c, ~ c \in [1,n] \}$ over the 
one-dimensional universe $[1,n\sigma]$, which is
roughly the inverse of the transform 
used in the previous section. We also store in an array $X[1,n]$ the pairs
$\langle r,rank_r(S,c) \rangle$, where $r=S[c]$, for the point corresponding 
to each column $c$ in the set. Those pairs are stored in row-major order in $X$,
that is, by increasing point value $(r-1) \cdot n + c$.

To query $rank_r(S,c)$ we compute the predecessor of $(r-1)\cdot n+c$, which
gives us its position $p$ in $X$.
If $X[p]$ is of the form $\langle r, v \rangle$, for some $v$, this means
that there are points in row $r$ and columns $[1,c]$ of the matrix, and thus
there are occurrences of $r$ in $S[1,c]$. Moreover, $v=rank_r(S,c)$ is the
value we must return. Otherwise, there are no points in row $r$ and columns
$[1,c]$ (i.e., our predecessor query returned a point from a 
previous row), and thus
there are no occurrences of $r$ in $S[1,c]$. Thus we return zero.

\paragraph*{Example} Fig.~\ref{fig:lbub} also illustrates the upper-bound
technique on string $S=bbcab$, of length $n=5$ over an alphabet of size
$\sigma=3$. It corresponds to the lower matrix in the figure, which is read
row-wise and the 1s are written as $n=5$ points in a universe of size
$n\sigma = 15$. To each point we associate the row it comes from and its
rank in the row, in array $X$. Now the query $rank_b(S,3)$ is converted into
query $pred(8) = 3$ ($8 = 5 \times 1 + 3$). This yields
$X[3]=\langle 2,2 \rangle$, the first 2 indicating that there are $b$s 
up to position 3 in $S$ ($b$ is the 2nd alphabet symbol), and the second 2 
indicating that there are 2 $b$s in the range,
so the answer is 2. Instead, a query like $rank_c(S,2)$ would be translated
into $pred(12)=4$ ($12 = 5 \times 2 + 2$). This yields $X[4]=\langle 2,3 
\rangle$. Since the first component is not 3, there are no $c$s up to
position 2 in $S$ and the answer is zero. \hfill $\Box$

\medskip

This solution requires $n\lg\sigma + n\lg n$ bits for the pairs of $X$, 
on top of the 
space of the predecessor structure. If $\sigma \le n$ we can reduce this extra 
space to $2n\lg\sigma$ by storing the pairs $\langle r,rank_r(S,c) \rangle$ in a
different way. We virtually cut the string into chunks of length $\sigma$, and 
store the pair as 
$\langle r, rank_r(S,c)-rank_r(S,c-(c~\mathrm{mod}~\sigma))\rangle$, that is,
we only store the number of occurrences of $c$ from the beginning of the current
chunk. Such a pair requires $2\lg\sigma$ bits. The rest of
the $rank_r$ information, that is, up to the beginning of the chunk, is 
obtained in constant time and $O(n)$ bits using the reduction to chunks of
\citeN{GMR06}: They store a bitmap 
$A[1,2n]$ where the matrix is traversed row-wise and we append to $A$ a 1 for
each 1 found in the matrix and a 0 each time we move to the next chunk (so
we append $n/\sigma$ 0s per row). Then the remaining information for 
$rank_r(S,c)$ is $rank_r(S, c - (c~\mathrm{mod}~\sigma)) = 
 select_0(A,p_1) - select_0(A,p_0)- \lfloor c/\sigma\rfloor$, where
 $p_0 = (r-1)\cdot n/\sigma$ is the number of chunks in previous rows and
$p_1 = p_0 + \lfloor c/\sigma\rfloor$ is the number of chunks preceding the
current one (we have simplified the formulas by assuming $\sigma$ divides $n$).
The $select_0(A,\cdot)$ operations map chunk numbers to positions in $A$, and
the final formula counts the number of 1s in between.

\begin{theorem}
Given a solution for predecessor search on a set of $n$ keys chosen from a
universe of size $u$, that occupies space $s(n,u)$ and answers in time $t(n,u)$,
there exists a solution for $rank$ queries on a sequence of length $n$ over 
an alphabet $[1,\sigma]$ that runs in time $t(n,n\sigma)+O(1)$ and occupies 
$s(n,n\sigma)+O(n\lg\sigma)$ bits.
\end{theorem}

In the extended version of their article, \citeN{PT08} 
give an upper bound matching the lower bound of branch 2 and using $O(n\lg u)$ 
bits for $n$ elements over a universe $[1,u]$. In Appendix~\ref{sec:pred} we 
show that the same time can be achieved with space $O(n\lg(u/n))$, which is not 
surprising (they have given hints, actually) but we opt for completeness. By 
using this predecessor data structure, the following result is immediate.

\begin{theorem}
A string $S[1,n]$ over alphabet $[1,\sigma]$ can be represented using
$O(n\lg\sigma)$ bits, so that operation $rank$ is solved in time 
$O\left(\lg \frac{\lg\sigma}{\lg w}\right)$.
\end{theorem}

Note that, within $O(n\lg\sigma)$ bits, operations $access$ and $select$ can 
also be solved in constant time: we can add a plain representation of $A$ to
have constant-time $access$, plus a succinct representation \cite{GMR06}
that supports constant-time $select$, adding $2n\lg\sigma+o(n\lg\sigma)$
bits in total.

When $\sigma > n$, we can add a perfect hash function mapping 
$[1,\sigma]$ to the (at most) $n$ symbols actually occurring in $S$, in 
constant time, and then $S[1,n]$ can be built over the mapped alphabet of 
size at most $n$. The hash function can be implemented as an array of
$n\lg\sigma$ bits listing the symbols that do appear in $S$ plus 
$O(n\lg\lg(\sigma/n))$ bits for a mmphf to map from $[1,\sigma]$ to the array
\cite{BBPV09}. Therefore, in this case we obtain the improved time
$O\left(\lg \frac{\lg n}{\lg w}\right)$.

\section{Using Succinct Space} \label{sec:succinct}

We design a sequence representation using $n\lg\sigma + o(n\lg\sigma)$
bits (i.e., succinct) that answers $access$ and $select$ queries in 
almost-constant time, and $rank$ in time $O\left(\lg\frac{\lg\sigma}{\lg w}\right)$. 
This is done in two phases: a constant-time solution for 
$\sigma = w^{O(1)}$, and then a solution for general alphabets.

\subsection{Succinct Representation for Small Alphabets}

Using multiary wavelet trees \cite{FMMN07,GRR08} we can obtain succinct space
and $O\left(1+\frac{\lg\sigma}{\lg\lg n}\right)$ time for $access$, $select$ and
$rank$. This is constant for $\lg\sigma = O(\lg\lg n)$. We start by extending
this result to the case $\lg\sigma = O(\lg w)$, as a base case for handling
larger alphabets thereafter. More precisely, we prove the following result.

\begin{theorem} \label{thm:upper-smallsuccinct}
A string $S[1,n]$ over alphabet $[1,\sigma]$, $\sigma \le n$, 
can be represented using
$n\lg\sigma + o(n)$ bits so that operations $access$,
$select$ and $rank$ can be solved in time $O\left(1+\frac{\lg\sigma}{\lg w}\right)$.
\end{theorem}

A multiary wavelet tree for $S[1,n]$ divides, at the root node $v$, the alphabet
$[1,\sigma]$ into $r$ contiguous regions of the same size. A sequence $R_v[1,n]$
recording the region each symbol belongs to is stored at the root node $v$ (note
$R_v$ is a sequence over alphabet $[1,r]$). This node has $r$ children, each 
handling the subsequence of $S$ formed by the symbols belonging to a given 
region. The children are decomposed recursively, thus the wavelet 
tree has height $h=\lceil\lg_r \sigma\rceil$. 
Queries $access$, $select$ and $rank$ on 
sequence $S[1,n]$ are carried out via $O(\lg_r \sigma)$ similar queries on 
the sequences $R_v$ stored at wavelet tree nodes \cite{GGV03}. By choosing $r$ 
such that $\lg r = \Theta(\lg\lg n)$, it turns out that the operations on the
sequences $R_v$ can be carried out in constant time, and thus the
cost of the operations on the original sequence $S$ is 
$O\left(1+\frac{\lg\sigma}{\lg\lg n}\right)$ \cite{FMMN07}.
\citeN{GRR08} show how to retain these time complexities within only
$n\lg\sigma+o(n)$ bits of space.

In order to achieve time $O\left(1+\frac{\lg\sigma}{\lg w}\right)$, we need to handle in
constant time the operations over alphabets of size $r = w^\beta$, for some
$0<\beta<1$, so that $\lg r = \Theta(\lg w)$. This time we cannot resort to 
universal tables of size $o(n)$, but rather must use bit manipulation on the 
RAM model. The description of bit-parallel operations is rather technical; 
readers interested only in the result (which is needed afterwards) can skip to 
Section~\ref{sec:rank-succinct}.

The sequence $R_v[1,n]$ is stored as the concatenation of $n$ fields of 
length $\ell=\lceil \lg r\rceil$, into consecutive machine words. Thus achieving constant-time 
$access$ is trivial: To access $R_v[i]$ we simply extract the corresponding
bits, from the $(1+(i-1)\cdot \ell)$-th to the $(i\cdot \ell)$-th, from
one or two consecutive machine words, using bit shifts and masking.

Operations $rank$ and $select$ are more complex.
We will proceed by cutting the sequence $R_v$ into blocks of length
$b=\Theta(w^\alpha/\ell)$ symbols, for some $\beta<\alpha<1$. First we show 
how, given a block number $i$ and a symbol $a$, we extract from $R[1,b] =
R_v[(i-1)\cdot b+1, i \cdot b]$ a bitmap that marks the values $R[j]=a$.
Then we use this result to achieve constant-time $rank$ queries.
Next, we show how to solve predecessor queries in constant time, for several
fields of length $\lg w$ bits fitting in a machine word. Finally,
we use this result to obtain constant-time $select$ queries. In the following,
we will sometimes write bit-vector constants; in those, bits are written
right-to-left, that is, the rightmost bit is that at the bitmap position 1.

\paragraph*{Projecting a block}
\label{sec:projecting}

Given sequence $R[1,b] = R_v[1+(i-1)\cdot b, i \cdot b]$, which is of bit
length $b\cdot\ell= \Theta(w^\alpha) = o(w)$,
and given $a \in [1,r]$, we extract $B[1,b\cdot\ell]$ such that 
$B[j\cdot\ell]=1$ iff $R[j]=a$.
To do so, we first compute $X = a \cdot (0^{\ell-1} 1)^b$.
This creates $b$ copies of $a$ within $\ell$-bit long fields. 
Second, we compute $Y = R ~\textsc{xor}~ X$, which will have zeroed fields
at the positions $j$ where $R[j]=a$. To identify those fields, we compute
$Z = (10^{\ell-1})^b - Y$, which will have a 1 at the highest bit of the
zeroed fields in $Y$. Finally, $B = Z ~\textsc{and}~ (10^{\ell-1})^b$
isolates those leading bits.

\paragraph*{Constant-time rank queries}
\label{sec:bitparrank}

We now describe how we can do rank queries in constant time for $R_v[1,n]$. 
Our solution follows that of \citeN{Mun96}.
We choose a superblock size $s = w^2$ and a block size $b = (\sqrt{w}-1)/\ell$.
For each $a \in [1,r]$, we store the accumulated values per superblock,
$rank_a(R_v,i \cdot s)$ for all $1 \le i \le n/s$. We also store the 
within-superblock accumulated values per block,
$rank_a(R_v,i \cdot b) - rank_a(R_v, \lfloor (i\cdot b) / s \rfloor \cdot s)$,
for $1 \le i \le n/b$.
Both arrays of counters require, over all symbols,
$r ((n/s)\cdot w + (n/b)\cdot \lg s) = O(n w^\beta (\lg w)^2/\sqrt{w})$ bits.
Added over the $O\left(\frac{\lg\sigma}{\lg w}\right)$ wavelet tree levels, the space
required is $O(n \lg \sigma \lg w/w^{1/2-\beta})$ bits. This is $o(n\lg\sigma)$ 
for any $\beta<1/2$.

To solve a query $rank_a(R_v,i)$, we need to add up three values: 
$(i)$ the superblock accumulator at position $\lfloor i/s \rfloor$, 
$(ii)$ the block accumulator at position $\lfloor i/b \rfloor$, 
$(iii)$, the bits set at $B[1,(i~\mathrm{mod}~ b)\cdot\ell]$, where
$B$ corresponds to the values equal to $a$ in 
$R_v[\lfloor i/b \rfloor \cdot b + 1,\lfloor i/b \rfloor \cdot b + b]$.
We have just shown how to extract $B[1,b\cdot\ell]$ from $R_v$,
so we count the number of bits set in 
$C = B ~\textsc{and}~ 1^{(i~\mathrm{mod}~ b)\cdot\ell}$.

This counting is known as a popcount operation. Given a bit block $C$ of length 
$b\ell=\sqrt{w}-1$, with bits possibly set at positions multiple of $\ell$, we 
popcount it using the following steps: 
\begin{enumerate}
\item We first duplicate the block $b$ times into $b$ fields. That is, we
compute $X = C \cdot (0^{b\ell-1}1)^b$.
\item We now isolate a different bit in each different field. This is done
with $Y=X~\textsc{and}~(0^{b\ell} 10^{\ell-1})^b$. This will isolate the $i$th
aligned bit in field $i$. 
%     xyzwxyzwxyzwxyzw = X b=4
%&00001000010000100001
%=0000x0000y0000z0000w = Y
\item We now sum up all those isolated bits using the multiplication 
$Z=Y\cdot (0^{b\ell+\ell-1} 1)^b$. 
%*00001000010000100001
%=    ^[x+y+z+w]
The result of the popcount operation lies at the bits 
$Z[b^2\ell,b^2\ell+\lg b-1]$.
\item We finally extract the result as 
$c=(Z\gg (b^2\ell-1))~\textsc{and}~(1^{\lg b})$. 
\end{enumerate}

%Note the result fits in the machine word because $b^2\ell+\lg b =
%(\sqrt{w}-1)^2/\ell + \lg (\sqrt{w}-1)/ell
%< w/\ell - 2sqrt{w}/beta lg w + 1/ell + 1/(2 beta) = O(w/\lg w).

\paragraph*{Constant-time select queries}
\label{sec:bitparsel}

We now describe how we can do $select$ queries in constant time for $R_v[1,n]$.
Our solution follows that of \citeN{Cla96}. For each $a \in [1,r]$, 
consider the virtual bitmap $B_a[1,n]$ so that $B_a[j]=1$ iff $R_v[j]=a$.
We choose a superblock size $s=w^2$ and a block size $b = w^{1/3}/(2\lg r)$. 
Superblocks
contain $s$ 1-bits and are of variable length. They are called {\em dense} if
their length is at most $w^4$, and {\em sparse} otherwise. We store all the
positions of the 1s in sparse superblocks, which requires $O(n/w)$ bits of
space as there are at most $n/w^4$ sparse superblocks. For dense superblocks 
we only store their starting position in $R_v$ and a pointer to a memory area. 
Both pointers require $O(n/w)$ bits since there are at most $n/w^2$ superblocks.

We divide the dense superblocks into blocks of $b$ 
1s. Blocks are called {\em dense} if their length is at most 
$w^{2/3}$, and {\em sparse} otherwise. We store all the positions of the 1s
in sparse blocks. Since each position requires only $\lg(w^4)$ as it is within 
a dense superblock, and there are at most $n/w^{2/3}$ sparse blocks, the total 
space for sparse blocks is $O((n/w^{2/3}) b \lg w) = O(n /w^{1/3})$
bits. For dense blocks we store only their starting position within their
dense superblock, which requires $O((n/b) \lg w) = O(n(\lg w)^2 /w^{1/3})$ bits.

The space, added over the $r$ symbols, is
$O(r n (\lg w)^2 / w^{1/3}) = O(n (\lg w)^2 / w^{1/3-\beta})$. Summing for 
$O\left(\frac{\lg\sigma}{\lg w}\right)$ wavelet tree levels, the total space is
$O(n \lg\sigma \lg w/ w^{1/3-\beta})$ bits. This is $o(n\lg\sigma)$ for any 
$\beta < 1/3$.

In order to compute a $select_a(R_v,j)$ query, we use the data structures for
virtual bitmap $B_a[1,n]$. If $\lfloor j/s \rfloor$ is a sparse superblock, 
then the answer is readily stored. If it is a dense superblock, we only know its
starting position and the offset $o = j - (j ~\mathrm{mod}~ s)$ of the
query within its superblock. Now, if $\lfloor o/b \rfloor$ is a sparse block
in its superblock, then the answer (which must be added to the starting
position of the superblock) is readily stored. If it is a dense block, we
only know its starting position in $R_v$ (and in $B_a$), but now we only have to complete
the search within an area of length $b=O(w^{1/3}/\lg w)$ in $B_a$. 
We have showed how to extract a chunk $B[1,b\cdot\ell]$
from $R_v$, so that $B[i\cdot\ell] = B_a[i]$. Now we detail how we complete a 
$select$ query within a chunk of length $b\cdot\ell = O(w^{1/3})$ for the 
remaining $j'=j - (j ~\mathrm{mod}~ b)$ bits.
This is based on doing about $w^{1/3}$ parallel popcount operations on about
$w^{1/3}$ bit blocks. We proceed as follows:

\begin{enumerate}
\item Duplicate $B$ into $b$ superfields with $X=B\cdot(0^{k-1}1)^b$, where
$k=2b^2\ell$ is the superfield size.
\item Compute $Y=X~\textsc{and}~(0^{k-b\ell}1^{b\ell})\ldots(0^{k-2\ell}1^{2\ell}) (0^{k-\ell}1^{\ell})$. 
This operation will keep only the first $i$ aligned bits in superfield $i$.
\item Do popcount in parallel on all superfields using the algorithm 
described in Section~\ref{sec:bitparrank}. Note that each superfield will have 
capacity $k=2b^2\ell$, but only the first $b\ell$ bits in it are set, and the
alignment is $\ell$. Thus the popcount operation will have enough available 
space in each superblock to operate. 
\item Let $Z$ contain all the partial counts for all the prefixes of $B$. 
We need the position in $Z$ of the first count equal to $j'$. We use the same
projecting method described in Section~\ref{sec:projecting} to spot the
superfields equal to $j'$ (the only difference is that superfields are much
wider than $\lg w$, namely of width $\ell=k$, but still all fits in a machine
word). This method returns a word $W[1,2b^3\ell]$ such that $W[k\cdot i]=1$ iff 
the $i$th superfield of $Z$ is equal to $j'$. 
\item Isolate the least significant bit of $W$ with 
$V = W~\textsc{and}~ (W ~\textsc{xor}~(W-1))$.
\item The final answer to $select_1(B,j')$ is the position of the only 1 in
$V$, divided by $k$. This is easily computed by using 
mmphfs over the set $\{ 2^{ki},~ 1 \le i \le b\}$. Existing 
data structures \cite{BBPV09} take constant time and $O(b \lg w) = O(w)$ bits.
Such a data structure is universal and requires the same space as systemwide
pointers.
\end{enumerate}

\paragraph*{Space analysis}

We choose $r =w^\beta$ to be a power of 2, $r=2^\ell$. This is always possible 
because it is equivalent to finding an integer $\ell = \beta\lg w$, where we 
can choose any constant $0<\beta<1/3$ and any $\ell=\Theta(\lg w)$ (e.g., one 
solution is $\beta = \lfloor \frac{\lg w}{4} \rfloor/\lg w$, 
$\ell=\lfloor \lg w/4\rfloor$, and $r = 2^{\lfloor\lg w/4\rfloor} 
\approx w^{1/4}$). 
In this case the wavelet tree simply stores, at level $l$, the bits 
$(l-1)\cdot\ell+1$ to $l\cdot\ell$ of the binary descriptions of the symbols
of $S$. 
The wavelet tree has height $h=\lceil \lg_r\sigma\rceil =
\lceil(\lg\sigma)/\ell\rceil$, so it will store sequences of symbols of
$\ell=\lg r$ bits in each of the $h$ levels except in the first, where it will 
store a sequence of symbols of $\lceil\lg\sigma\rceil-(h-1)\ell \le \ell$ bits.
%(h-1)l >= lg s - l
%ceil(lg s) - (h-1)l <= ceil(lg s) - lg s + l <= l pues es entero
The total adds up to $n \lceil \lg\sigma \rceil$ bits.

This is not fully satisfactory when $\sigma$ is not a power of two. In
this case we proceed as follows. We choose an integer $y = \lg\sigma -
\Theta(\lg\lg n)$ as the number of bits of the representation that will be
stored integrally, just as explained. The other $x = \lg\sigma - y$ bits
(where $x$ is not an integer) will be represented as symbols over alphabet
$[1,\sigma_0]=[1,\lceil 2^x\rceil] = [1,\lceil\sigma/2^y\rceil]$.
By construction, $\sigma_0 = \lg^{\Theta(1)} n$, thus we can represent the
sequence of $x$ highest bits (i.e., the numbers $\lceil S[i]/2^y\rceil$)
using the space-efficient wavelet tree of \citeN{GRR08}.
This will take $n\lg\sigma_0+o(n)$ bits and support $access$, $select$ and
$rank$ in constant time, and will act as the root level of our whole
wavelet tree. For each value $c \in [1,\sigma_0]$ we will store, as a child
of that root,
a separate wavelet tree handling the subsequence of positions $i$ such
that $\lceil S[i]/2^y\rceil = c$. These wavelet trees will handle the
$y$ lower bits of the sequence with the technique of the previous paragraph,
which will take $ny$ bits and solve the three queries in 
$O\left(1+\frac{y}{\lg w}\right)$ time. Adding up the spaces we get
$n\lg\sigma_0 + ny + o(n) < n (\lg (1+\sigma/2^y) + y + o(1))
= n (\lg(\sigma+2^y) + o(1)) = n(\lg(\sigma(1+1/\lg^{\Theta(1)}n))+o(1))
\le n(\lg\sigma + 1/\lg^{\Theta(1)}n+o(1)) = n\lg\sigma+o(n)$.

To this space we must add the $o(n\lg\sigma)$ bits of the extra structures
to support $rank$ and $select$ on the wavelet tree levels.
The special level using less than $\ell$ bits can use the same $\alpha$ value
of the next levels without trouble (actually the redundancy may be lower
since more symbols can be packed in the blocks).

In order to further reduce the redundancy to $o(n)$ bits, we use the scheme we
have described only for $w > \lg^d n$, for some constant $d$ to be defined soon.
For smaller $w$, we directly use the scheme of \citeN{GRR08}, which uses 
$n\lg\sigma+o(n)$ bits and solves all the operations in time 
$O\left(1+\frac{\lg\sigma}{\lg\lg n}\right)=O\left(1+\frac{\lg\sigma}{\lg w}\right)$.
For the larger $w$ case, and choosing our example $\beta \le 1/4$,
our redundancy is of the form $O(n\lg\sigma \cdot (\lg w/w^{1/3-\beta})) = 
O(n \lg n \cdot (d\lg\lg n / (\lg n)^{d/12}))$, which is made $o(n)$ by choosing
any $d> 12$ (a smaller $d$ can be chosen if a smaller $\beta$ is used).

Finally, we have the space redundancy of the wavelet tree pointers. On binary
wavelet trees this is easily solved by concatenating all the bitmaps 
\cite{MN07}. This technique can be extended to $r$-ary wavelet trees, but in
this case a simpler solution is as follows. As the wavelet tree has 
a perfect $r$-ary structure, we deploy its nodes levelwise in memory. For
each level, we concatenate all the sequences of the nodes, read left-to-right,
into a large sequence of at most $n$ symbols. Then the node position we want
at each level can be algebraically computed from that of the previous or next
level,
whereas its starting positions in the concatenation of sequences can be marked 
in a bitmap of length $n$, which will have at most $r^j$ 1s for the level $j$ 
of the wavelet tree. Using the representation of \citeN{RRR07} for this bitmap,
the space is $O(r^j \lg(n/r^j))+o(n)$ bits. Thus the space is
dominated by the last level, which has $r^j = \sigma/w^\beta \le n/w^\beta$ 
1s, giving overall
space $O(n\lg w /w^\beta)+o(n)=o(n)$ bits. Then any pointer can be retrieved 
with a constant-time $select$ operation on the bitmap of its level.

\subsection{Succinct Representation for Larger Alphabets} 
\label{sec:rank-succinct}

We now assume $\lg\sigma = \omega(\lg w)$ and develop fast succinct solutions 
for these larger alphabets.
We build on the solution of \citeN{GMR06}. They first cut $S$ into
chunks of length $\sigma$. With the bitvector $A[1,2n]$ described in 
Section~\ref{sec:upper} they reduce all the queries, in constant time, to 
within a chunk. For each chunk they store a bitmap $X[1,2\sigma]$ where the 
number of occurrences of each symbol $a \in [1,\sigma]$ in the chunk, $n_a$, is 
concatenated in unary, $X = 1^{n_1}0 1^{n_2}0 \ldots 1^{n_\sigma} 0$. 
Now they introduce two complementary solutions. 

\paragraph*{Constant-time select}

The first one stores, for each consecutive symbol $a \in [1,\sigma]$, the 
chunk positions where it appears, in increasing order. 
Let $\pi$ be the resulting permutation, which is stored with the
representation of \citeN{MRRR03}.
This requires $\sigma\lg\sigma (1+1/f(n,\sigma))$ bits and computes any 
$\pi(i)$ in constant time and any $\pi^{-1}(j)$ in time $O(f(n,\sigma))$,
for any $f(n,\sigma) \ge 1$. With this representation they solve, within the 
chunk, $select_a(i) = \pi(select_0(X,a-1)-(a-1)+i)$ in constant time and
$access(i) = 1+rank_0(select_1(X,\pi^{-1}(i)))$ in time $O(f(n,\sigma))$.

For $rank_a(i)$, they basically carry out a predecessor search within the 
interval of $\pi$ that corresponds to $a$: 
$[select_0(X,a-1)-(a-1)+1,select_0(X,a)-a]$. They have a sampled
predecessor structure with one value out of $\lg\sigma$, which takes just
$O(\sigma)$ bits. With this structure they reduce the interval to size
$\lg\sigma$, and a binary search completes the process, within overall
time $O(\lg\lg\sigma)$.

To achieve optimal time, we sample one value out of 
$\frac{\lg\sigma}{\lg w}$. We build the predecessor data structures of
\citeN{PT08} mentioned in Section~\ref{sec:upper}.
Over all the symbols of the chunk,
these structures take $O\left(\left(n/\frac{\lg\sigma}{\lg w}\right) \lg \sigma\right) =
O(n \lg w) = o(n \lg \sigma)$ bits (as we assumed
$\lg\sigma=\omega(\lg w)$). The predecessor structures take time
$O\left(\lg\frac{\lg\sigma}{\lg w}\right)$ (see Theorem~\ref{thm:smallpred} in 
Appendix~\ref{sec:pred}). The final binary search time also takes time
$O\left(\lg\frac{\lg\sigma}{\lg w}\right)$.

\paragraph*{Constant-time access}

This time we use the structure of Munro et al.\ on $\pi^{-1}$, so we compute
any $\pi^{-1}(j)$ in constant time and any $\pi(i)$ in time $O(f(n,\sigma))$.
Thus we get $access$ in constant time and $select$ in time $O(f(n,\sigma))$.

Now the binary search of $rank$ needs to compute values of $\pi$, which is 
not anymore constant time. This is why \citeN{GMR06} obtained
time slightly over $\lg\lg\sigma$ time for $rank$ in this case. We instead set 
the sampling step to $\left(\frac{\lg \sigma}{\lg w}\right)^\frac{1}{f(n,\sigma)}$. The 
predecessor structures on the sampled values still answer in time 
$O\left(\lg\frac{\lg\sigma}{\lg w}\right)$, but they take 
$O\left(\left(n/\left(\frac{\lg\sigma}{\lg w}\right)^\frac{1}{f(n,\sigma)}\right)\lg\sigma\right)$ bits of 
space.  This is
$o(n\lg\sigma)$ provided $f(n,\sigma) = o\left(\lg\frac{\lg\sigma}{\lg w}\right)$.
On the other hand, the time for the binary search is
$O\left(\frac{f(n,\sigma)}{f(n,\sigma)}\lg\frac{\lg\sigma}{\lg w}\right)$, as desired.

\medskip

The following theorem, which improves upon the result of 
\citeN{GMR06} (not only as a consequence of a higher low-order space term),
summarizes our result. Note that we do not mention the limit
$f(n,\sigma) = o\left(\lg\frac{\lg\sigma}{\lg w}\right)$, as if a larger $f(n,\sigma)$ is
desired we can always use a smaller one (and be faster). We also omit the
condition $\lg\sigma = \omega(\lg w)$ because otherwise the result also holds
by Theorem~\ref{thm:upper-smallsuccinct}.

\begin{theorem} \label{thm:upper-succinct}
A string $S[1,n]$ over alphabet $[1,\sigma]$, $\sigma \le n$, can be 
represented using
$n\lg\sigma + o(n\lg\sigma)$ bits, so that, given any function
$f(n,\sigma) = \omega(1)$,
$(i)$ operations $access$ and $select$ can be solved in time $O(1)$ and
$O(f(n,\sigma))$, or vice versa, and $(ii)$ $rank$ can be solved 
in time $O\left(\lg \frac{\lg\sigma}{\lg w}\right)$.
\end{theorem}

Note that we can partition into chunks only of $\sigma \le n$. If $\sigma =
o(n\lg n)$ we can still apply the same scheme using a single chunk, and the space
overhead for having $\sigma > n$ will be $O(\sigma) = o(n\lg\sigma)$. For larger
$\sigma$, however, we must use a mechanism like the one used at
the end of Section~\ref{sec:upper}, mapping $[1,\sigma]$ to $[1,n]$. However, 
this adds at least $n\lg\sigma$ bits to the space, and thus the space is
not succinct anymore, unless $\sigma$ is much larger,
$\lg\sigma = \omega(\lg n)$, so that the space of the mapping array dominates.
For simplicity we will consider only the case $\sigma \le n$ in the rest
of the article.

\section{Compressing the Space} \label{sec:compressed}

Now we compress the space of the succinct solutions of the previous sections.
First we achieve zeroth-order compression (of the data and the
redundancy) by using an existing compression booster \cite{BCGNN12}. Second,
we reach high-order compression by designing an index that operates over
a compressed representation \cite{FV07} and simulates the working of a
succinct data structure of the previous section.

\subsection{Zero-order Compression}

\citeN[Thm.~2]{BCGNN12} showed how, given a sequence representation 
$\mathcal{R}$ using $n\lg\sigma (1+ r(n,\lg\sigma))+o(n)$ bits, where 
$r(n,\lg\sigma) = O(1)$ is nonincreasing with $\sigma$,
its times for $access$, $select$ and $rank$ can be maintained while reducing 
its space to $nH_0(S) (1+r(n,\Theta(\lg\lg n))+o(n)$ bits.%
\footnote{They used the case $r(n,\lg\sigma)=1/\lg\lg\sigma$, but their 
derivation is general.} This can be done even if 
$\mathcal{R}$ works only for $\sigma \ge \lg^c n$ for some constant $c$.

The technique separates the symbols according to their frequencies
into $\lg^2 n$ classes. The sequence of classes is represented using a
multiary wavelet tree \cite{FMMN07}, and the subsequences of the symbols of 
each class are represented with an instance of $\mathcal{R}$ if the local
alphabet size is $\sigma' \ge \lg^c n$, or with a multiary wavelet tree
otherwise. Hence the global per-bit redundancy can be upper bounded by
$r(n,c\lg\lg n)$ and it is shown that the total number of bits represented is
$nH_0(S)+O(n/\lg n)$.

We can use this technique to compress the space of our succinct
representations. By using Theorem~\ref{thm:upper-smallsuccinct} as our
structure $\mathcal{R}$, where we can use $r(n,\lg\sigma)=0$,
we improve upon \citeN{FMMN07} and \citeN{GRR08}. 

\begin{theorem} \label{thm:upper-smallH0}
A string $S[1,n]$ over alphabet $[1,\sigma]$, $\sigma\le n$, can be 
represented using $nH_0(S)+o(n)$ bits so that operations $access$,
$select$ and $rank$ can be solved in time $O\left(1+\frac{\lg\sigma}{\lg w}\right)$.
\end{theorem}

To obtain better times when $\lg\sigma = \omega(\lg w)$, we
use Theorem~\ref{thm:upper-succinct} as our structure $\mathcal{R}$.
A technical problem is that \citeN{BCGNN12} apply $\mathcal{R}$
over smaller alphabets $[1,\sigma']$, and thus in
Theorem~\ref{thm:upper-succinct} we would sample one position out of 
$\frac{\lg\sigma'}{\lg w}$, obtaining 
$O\left(\lg\frac{\lg\sigma'}{\lg w}\right)$ time and 
$O\left(n\lg\sigma' \frac{\lg w}{\lg \sigma'}\right)$ bits of space, which
is $o(n\lg\sigma')$ only if $\lg\sigma' = \omega(\lg w)$ (this is why we have
used Theorem~\ref{thm:upper-succinct} only in that case).
To handle this problem, we will use a sampling
of size $\frac{\lg\sigma}{\lg w}$ 
(or $\left(\frac{\lg\sigma}{\lg w}\right)^\frac{1}{f(n,\sigma)}$ in the case of
constant-time $access$), even if the alphabet of the local sequence is of size
$\sigma'$. As a consequence, the redundancy will be 
$O(n\lg\sigma'\,\frac{\lg w}{\lg\sigma}) = o(n\lg\sigma')$ and the time for
$rank$ will stay $O\left(\lg\frac{\lg\sigma}{\lg w}\right)$ (instead of
$O\left(\lg\frac{\lg\sigma'}{\lg w}\right)$). Similarly, we always use sampling
rate $f(n,\sigma)$ instead of $f(n,\sigma')$. Therefore our redundancy is
$r(n,\lg\sigma) = O\left(\frac{\lg w}{\lg\sigma} + \frac{1}{f(n,\sigma)}\right)$, which is $o(1)$ if
$\lg\sigma=\omega(\lg w)$.

Still, in the first levels where $\sigma'=O(1)$, the redundancy of
Theorem~\ref{thm:upper-succinct} contains space terms of the form $O(n)$ that
would not be $o(n\lg\sigma')$. To avoid this, we will use 
Theorem~\ref{thm:upper-smallsuccinct} up to $\frac{\lg\sigma'}{\lg w} \le 1$, 
where all times are constant, and the variant just described for larger
$\sigma'$. The result is an improvement over \citeN{BCGNN12}
(again, we do not mention the condition $\lg\sigma = \omega(\lg w)$ because
otherwise the result holds anyway by Theorem~\ref{thm:upper-smallH0}).

\begin{theorem} \label{thm:upper-H0}
A string $S[1,n]$ over alphabet $[1,\sigma]$, $\sigma \le n$, can be 
represented using $nH_0(S) + o(nH_0(S))+o(n)$ bits, so that, given any function
$f(n,\sigma)=\omega(1)$,
$(i)$ operations $access$ and $select$ can be solved in time $O(1)$ and 
$O(f(n,\sigma))$, or vice versa, and $(ii)$ $rank$ can be solved in time 
$O\left(\lg \frac{\lg\sigma}{\lg w}\right)$.
\end{theorem}

\subsection{Self-Indexing}

Likewise, we can improve upon the result of \citeANP{BCGNN12} that plugs a 
zeroth-order compressed sequence representation to obtain a $k$-th order 
compressed full-text self-index \cite[Thm.~5]{BCGNN12}. This result is not
subsumed by that of \citeN{BN11} because their index,
although obtaining better times, uses $O(n)$ extra bits of space. Ours is the
best result using only $o(n)(H_k(S)+1)$ bits of redundancy. We start with a 
version for small alphabets.

\begin{theorem}
Let $S[1,n]$ be a string over alphabet $[1,\sigma]$, $\lg\sigma =O(\lg w)$. 
Then we can represent $S$ using $nH_k(S) + o(n)$ bits, for any 
$k \le (\delta \lg_\sigma n) - 1$ and constant $0 < \delta < 1$, while 
supporting the following queries, for any function $f(n) = \omega(1)$: 
$(i)$ count the number of occurrences of a pattern $P[1,m]$ in $S$, in time 
$O(m)$; $(ii)$ locate any such occurrence in time $O(f(n)\lg n)$; 
$(iii)$ extract $S[l,r]$ in time $O(r-l+f(n)\lg n)$.
\end{theorem}

To obtain this result, we follow the proof of Theorem 5 of \citeN{BCGNN12}. 
Our zeroth-order compressed structure will be that of our
Theorem~\ref{thm:upper-smallH0},
with constant time for all the operations and space overhead 
$O(n/\lg^\gamma n)=o(n)$ bits, for some $0<\gamma<1$.
For operations $(ii)$ and $(iii)$, we sample one text position out of
$O(f(n)\lg n)$ in the suffix array to obtain the claimed times.

On general alphabets, we obtain the following result, where once again we
only need to prove the case $\lg\sigma = \omega(\lg w)$.

\begin{theorem}
Let $S[1,n]$ be a string over alphabet $[1,\sigma]$, $\sigma \le n$. Then we 
can represent $S$ using $nH_k(S) + o(n)(H_k(S)+1)$ bits, for any 
$k \le (\delta \lg_\sigma n) - 1$ and constant $0 < \delta < 1$, while 
supporting the following queries, for any $f(n,\sigma)=\omega(1)$: 
$(i)$ count the number of occurrences of a 
pattern $P[1,m]$ in $S$, in time $O\left(m \lg\frac{\lg \sigma}{\lg w}\right)$; 
$(ii)$ locate any such occurrence in time $O(f(n,\sigma)\lg n)$; 
$(iii)$ extract $S[l,r]$ in time $O(r-l+f(n,\sigma)\lg n)$.
\end{theorem}

Again we follow the proof of Theorem 5 of 
\citeN{BCGNN12}. First, if $f(n,\sigma) = 
\omega\left(\lg\frac{\lg \sigma}{\lg w}\right)$, we set it to
$f(n,\sigma)=\lg\frac{\lg \sigma}{\lg w}$,
to ensure that no operation will be slower than $rank$.
Our string structure will be that of
Theorem~\ref{thm:upper-H0} with constant-time $select$, $O(f(n,\sigma))$
time $access$, and $O(nH_0(S)(\lg w/\lg\sigma + 1/f(n,\lg\sigma))+o(n)$
bits of overhead. 
\citeANP{BCGNN12} partition the text into strings $S^i$, which are represented
to their zeroth-order entropy. The main issue is to upper bound the sum of
the redundancies over all the strings $S^i$ in terms of the total length $n$. 
More precisely, we need to bound the factor multiplying $|S^i|H_0(S^i)$,
$O(\lg w/\lg\sigma + 1/f(|S^i|,\lg\sigma))$, in terms
of $n$ and not $|S^i|$. However, we can simply use the sampling value 
$f(n,\sigma)$ for all the strings $S^i$ that are represented using
Theorem~\ref{thm:upper-H0}, regardless of the length $|S^i|$. Then their
Theorem~5 can be applied immediatly.

For operations $(ii)$ and $(iii)$, we again sample one out of 
$O(f(n,\sigma)\lg n)$ text positions in the suffix array, but instead of 
moving backward in the text using $rank$ and $access$, we move forward using 
$select$, as in \citeN[Sec.~4]{BN11}, which is constant-time.

\subsection{High-order Compression}

\citeN{FV07} showed how a string $S[1,n]$ over
alphabet $[1,\sigma]$ can be stored within $nH_k(S)+o(n\lg\sigma)$ bits,
for any $k=o(\lg_\sigma n)$, so that it offers constant-time $access$ to
any $O(\lg_\sigma n)$ consecutive symbols.

We provide $select$ and $rank$ functionality on top of this representation by
adding extra data structures that take $o(n\lg\sigma)$ bits, whenever
$\lg\sigma = \omega(\lg w)$. The technique is similar to those used by
\citeN{BHMR11} and \citeN{GOR10}, and we use the
terminology of Section~\ref{sec:rank-succinct}.
We divide the text logically into chunks, as with
\citeN{GMR06}, and for each chunk we store a 
mmphf $f_a$ for each $a \in [1,\sigma]$. Each $f_a$ stores the
positions where symbol $a$ occurs in the chunk, so that given the position
$i$ of an occurrence of $a$, $f_a(i)$ gives $rank_a(i)$ within the chunk.
All the mmphfs can be stored within $O(n\lg\lg\sigma) = o(n\lg\sigma)$
bits and can be queried in constant time \cite{BBPV09}.
With array $X$ we can know, given $a$, how many symbols smaller than $a$ are
there in the chunk.

Now we have sufficient ingredients to compute $\pi^{-1}$ in constant time:
Let $a$ be the $i$th symbol in the chunk (obtained in constant time using
Ferragina and Venturini's structure), then
$\pi^{-1}(i) = f_a(i) + select_0(X,a-1) - (a-1)$.
Now we can compute $select$ and $rank$ just as done in the ``constant-time
$access$'' branch of Section~\ref{sec:rank-succinct}. The resulting theorem
improves upon the results of \citeN{BHMR11} (they did not use mmphfs). 

\begin{theorem} \label{thm:upper-Hk}
A string $S[1,n]$ over alphabet $[1,\sigma]$, for $\sigma \le n$ and
$\lg\sigma = \omega(\lg w)$, can be represented using 
$nH_k(S) + o(n\lg\sigma)$ bits for any $k=o(\lg_\sigma n)$ so that, 
given any function
$f(n,\sigma) = \omega(1)$,
$(i)$ operation $access$ can be solved in constant time,
$(ii)$ operation $select$ can be solved in time $O(f(n,\sigma))$, 
and $(ii)$ operation $rank$ can be solved in time 
$O\left(\lg \frac{\lg\sigma}{\lg w}\right)$.
\end{theorem}

%very similar to the way Grossi Orlandi and Raman use mmphf in theorem 5
%(pages 12-13) in extended version of their paper:
%http://arxiv.org/abs/1006.5354 [arxiv.org]

To compare with the corresponding result by \citeN{GOR10}, who do
use mmphfs to achieve $nH_k(S)+O(n\lg\sigma/\lg\lg\sigma)$ bits, $O(1)$ time for
$access$ and $O(\lg\lg\sigma)$ time for $select$ and $rank$, we can fix 
$f(n,\sigma)=\lg\lg\sigma$ to obtain the same redundancy. Then we obtain the
same time for operations $access$ and $select$, and improved time for $rank$.
Their results, however, hold for any alphabet size, which we do not cover 
for the case $\lg\sigma = O(\lg w)$. We can, however, improve that branch
too, by using any superconstant sampling $g(n,\sigma)^\frac{1}{f(n,\sigma)}$,
for $\lg g(n,\sigma) = \omega(f(n,\sigma))$. Then the time for $rank$
becomes $O(\frac{f(n,\sigma)}{f(n,\sigma)}\lg g(n,\sigma))$. 
By using, say, $\lg g(n,\sigma)=f(n,\sigma)^2$, we get the following result.

\begin{theorem} \label{thm:lower-Hk}
A string $S[1,n]$ over alphabet $[1,\sigma]$, for $\lg\sigma = O(\lg w)$, 
can be represented using $nH_k(S) + o(n\lg\sigma)$ bits for any 
$k=o(\lg_\sigma n)$ so that, given any function
$f(n,\sigma) = \omega(1)$,
$(i)$ operation $access$ can be solved in constant time,
$(ii)$ operation $select$ can be solved in time $O(f(n,\sigma))$, 
and $(ii)$ operation $rank$ can be solved in time 
$O(f^2(n,\sigma))$.
\end{theorem}

This result, while improving that of \citeANP{GOR10}, is not necessarily optimal,
as no lower bound prevents us from reaching constant time for all the 
operations. We can achieve time optimality and $k$th order compression for 
small alphabet sizes, as follows. We build on the representation of
\citeN{FV07}. For $k=o(\lg_\sigma n)$, they 
partition the sequence $S[1,n]$ into chunks of $s = \frac{1}{2}\lg_\sigma n
= \omega(k)$ symbols, and encode the sequence of chunks $S'[1,n/s]$ over 
alphabet $[1,\sigma^s]=[1,\sqrt{n}]$ into zeroth-order entropy. This gives $k$th 
order compression of $S$ and supports constant-time access to any chunk. Now 
we add, for each $c \in [1,\sigma]$, a bitmap $B_c[1,n/s]$ 
so that $B_c[i]=1$ iff chunk $S'[i]$ contains an occurrence of symbol 
$c$. We store in addition a bitmap $C_c$ with the number of 
occurrences, in unary, of $c$ in all the chunks $i$ where $B_c[i]=1$.
That is, for each $B_c[i]=1$, we append $0^{m-1}1$ to $C_c$, where 
$m$ is the number of times $c$ occurs in the chunk $S'[i]$. Then we can
easily know the number of occurrences of any $c$ in $S'[1,i]$ using
$select_1(C_c,rank_1(B_c,i))$. With a universal table on the chunks,
of size $\sigma^{s+1}\lg s = O(\sqrt{n}\,\mathrm{polylog}(n)) = o(n)$, we can 
complete
the computation of any $rank_c(S,i)$ in constant time. Similarly, we can
determine in which chunk is the $j$th occurrence of any $c$ in $S'$, by
computing $select_1(B_c,1+rank(C_c,j))$, and then we can easily complete the
calculation of any $select_c(S,j)$ with a similar universal table, all in
constant time.

Let us consider space now. The $B_c$ bitmaps add up to $\sigma n/s$ bits, of
which at most $n$ are set. By using the representation of \citeN{RRR07}
we get total space $n\lg\frac{\sigma}{s} + 
O\left(n+\frac{(\sigma n/s)\lg\lg(\sigma n/s)}{\lg (\sigma n/s)}\right)$ bits, 
which is $o(n\lg\sigma)$ for any 
$\sigma = O(\lg^{1+o(1)}n)$
%$\sigma = O(\lg_\sigma n\,\sigma^{o(1)})$
and $\sigma=\omega(1)$.
On the other hand, the $C_c$ bitmaps add up to length $n$ and require 
$o(n\lg\sigma)$ bits of space for any $\sigma=\omega(1)$.

For constant $\sigma$, instead, we can represent the $B_c$ bitmaps in plain
form, using $O(\sigma n/s) = o(n)$ bits, and the $C_c$ bitmaps using 
Raman et al., as they have only $\sigma n/s = O(n/s)$ 1s, and thus their 
total space is $O(\frac{n\lg s}{s})+o(n) = o(n)$ bits. The same time 
complexities are maintained.

\begin{theorem} \label{thm:lower-Hk-constant}
A string $S[1,n]$ over alphabet $[1,\sigma]$, for 
$\sigma = O\left(\lg^{1+o(1)} n\right)$, 
can be represented using $nH_k(S) + o(n\lg\sigma)$ bits for any 
$k=o(\lg_\sigma n)$ so that operations $access$, $select$ and $rank$ can be
solved in constant time.
\end{theorem}

\section{Conclusions} \label{sec:concl}

This work considerably reduces the gap between upper and lower bounds for
sequence representations providing $access$, $select$ and $rank$ queries.
Most notably, we give matching lower and upper bounds
$\Theta\left(\lg\frac{\lg\sigma}{\lg w}\right)$ for operation $rank$,
which was the least developed one in terms of lower bounds. The issue of 
the space related to this complexity is basically solved as well: we
have shown it can be achieved even within compressed space,
and it cannot be surpassed within space $O(n\cdot w^{O(1)})$. On the other hand,
operations $access$ and $select$ can be solved, within the same compressed
space, in almost constant time (i.e., one taking $O(1)$ and the other as close
to $O(1)$ as desired but not both reaching it, unless we double the space).
Our new compressed representations improve upon most of the previous work.

There are still, however, some intriguing issues that remain unclear, which 
prevent us from considering this problem completely closed:

\begin{enumerate}
\item 
The lower bounds of \citeN{Gol09} leave open the door to achieving
constant time for $access$ and $select$ simultaneously, with 
$O(n\lg\sigma\,\frac{\lg\sigma}{w})$ bits of redundancy. 
That is, both could be constant time
with $o(n\lg\sigma)$ redundancy in the interesting case $\lg\sigma = o(w)$. 
We have achieved this when $\lg\sigma = O(\lg w)$, but it is open whether this
is possible in the area $\omega(\lg w) = \lg\sigma = o(w)$. In our solution,
this would imply computing $\pi$ and $\pi^{-1}$ in constant time on a 
permutation using $n\lg n + o(n\lg n)$ bits. A lower bound on the redundancy 
of permutations in the same paper \cite{Gol09}, 
$\Omega\left(n\lg n\,\frac{\lg n}{w}\right)$ bits, forbids this for 
$\lg n = \Theta(w)$ but not for $\lg n = o(w)$. It is an interesting open
challenge to achieve this or prove that a stronger lower bound holds.
\item 
While we can achieve constant-time $select$ and almost-constant time
for $access$ (or vice versa), only the second combination is possible within
high-order entropy space. Lower bounds on the indexing model
\cite{GOR10} show that this must be the case (at least in the general case
where $\lg\sigma=\Theta(w)$) as long as our solution builds on
a compressed representation of $S$ supporting constant-time access, as it has 
been the norm \cite{BHMR11,BCGNN12,GOR10}. Yet, it is not clear that this is the
only way to reach high-order compression.
\item 
We have achieved high-order compression with almost-constant
$access$ and $select$ times, and optimal $rank$ time, but on alphabets
of size superpolynomial in $w$. For smaller alphabets, although constant time 
seems to be possible, we achieved it only for 
$\sigma=O(\lg^{1+o(1)} n)$.
This leaves open the interesting band of alphabet sizes $\lg^{1+\Omega(1)}n = \sigma = w^{O(1)}$, where we have achieved only (any) superconstant time.
It is also unclear whether we can obtain $o(n)$ redundancy, instead of 
$o(n\lg\sigma)$, for alphabets polynomial in $w$, with high-order compression.
\end{enumerate}

\bibliographystyle{esub2acm}
\newcommand{\bibsc}[1]{\textsc{#1}}
\newcommand{\bibyear}[1]{#1}
\newcommand{\bibemphic}[1]{\textit{#1}}
\newcommand{\bibemph}[1]{\textit{#1}}
\bibliography{paper}

\appendix 

\section{Upper Bound for Predecessor Search}
\label{sec:pred}

We describe a data structure that stores a set $S$ of $n$ elements from universe 
$U=[1,u]$ in $O(n\lg(u/n))$ bits of space, while supporting predecessor queries 
in time $O(\lg\frac{\lg u-\lg n}{\lg w})$. 
We first start with a solution that uses $O(n\lg u)$ bits of space. 
We use a variant of the traditional recursive van Emde Boas solution
\cite{PT08}. Let $\ell \ge \lg u$ be the length of the keys. We choose $\ell$ as
the smallest value of the form $\ell=(\lg w-1)\cdot 2^i \ge \lg u$, for some 
integer $i\geq 0$ (note $\ell \le 2\lg u$).
We denote the predecessor data structure that stores a set $S$ 
of keys of length $\ell$ by $D^\ell(S)$. Given an element $x$ the predecessor 
data structure should return a pair $(y,r)$ where $y$ is the predecessor of $x$
in $S$ (i.e., the maximum value $\le x$ in $S$) and $r$ is the rank of $y$ in 
$S$ (i.e., the number of elements of $S$ 
smaller than or equal to $y$). If the key $x$ has no predecessor in $S$ (i.e., 
it is smaller than any key in $S$), the query should return $(0,0)$. 

We now describe the solution. We partition the set $S$ according to the most 
significant $\ell/2$ bits. We call
$h(x)$ the $\ell/2$ most significant bits of $x$, and
$l(x)$ the $\ell/2$ least significant bits of $x$, $x = 2^{\ell/2}h(x)+l(x)$.

Let $S_p = \{ x \in S, h(x)=p \}$ denote the set of all the elements $x$ such 
that $h(x)=p$, let $S'_p$ denote the set $S_p$ deprived of its minimal and
maximal elements, and let $\hat S_p = \{ l(x), x \in S'_p \}$ denote the 
set of lower parts of elements in $S'_p$. Furthermore, let 
$P = \{ h(x), x \in S\}$ denote the set of all distinct values of
$h(x)$ in $S$. The data structure consists of the following 
components:

\begin{enumerate}
\item A predecessor data structure $D^{\ell/2}(P)$.
\item A predecessor data structure $D^{\ell/2}(\hat S_p)$ for each $p\in P$
where $\hat S_p$ is non-empty. 
\item A dictionary $I(P)$ (a perfect hash function with constant time and linear space) that stores the set $P$. To each element $p\in P$, the dictionary associates the tuple $(m,r_m,M,r_M,q)$ with $m$ (respectively $M$) being the smallest (respectively largest) element in $S_p$, $r_m$ (respectively $r_M$) being the rank of $m$ (respectively $M$) in $S$, and $q$ a pointer to $D^{\ell/2}(\hat S_p)$. 
\end{enumerate}

We have described the recursive data structure. The base case is a predecessor 
data structure $D^{\lg w-1}(S)$ for a set $S$ of size $t$. Note that the set 
$S$ is a subset of $U=[1,2^{\lg w-1}]=[1,w/2]$. This structure is technical and
is described in Section~\ref{sec:bitparpred}. It encodes $S$ using $O(t\lg |U|)
= O(t \lg w)$ bits and answers predecessor queries in constant time.

We now get back to the main data structure and describe how queries are done
on it. Given a key $x$, we first query $I(P)$ for the key $p=h(x)$. Now,
depending on the result, we have two cases:
\begin{enumerate}
\item The dictionary does not find $p$. Then we query $D^{\ell/2}(P)$ for the key $p-1$. This returns a pair $(y,r)$. If $(y,r)=(0,0)$ we return $(0,0)$. Else we search $I(P)$ for $y$, which returns a tuple $(m,r_m,M,r_M,q)$, and the final answer is $(M,r_M)$.
\item The dictionary finds $p$ and returns a tuple $(m,r_m,M,r_M,q)$. We have the following subcases:
\begin{enumerate} 
\item We have $x<m$. Then we proceed exactly as in case 1.
\item We have $x=m$, then the answer is $(m,r_m)$.
\item We have $x\geq M$, then the answer is $(M,r_M)$.
\item We have $m<x<M$. Then we query $D^{\ell/2}(\hat S_p)$ (pointed by $q$) for the key $l(x)$. This returns a tuple $(y,r)$. The final answer answer is 
$(2^{\ell/2}\,p+y,r_m+r)$ if $(y,r)\neq (0,0)$ and $(m,r_m)$ otherwise.
\end{enumerate}
\end{enumerate}

\paragraph*{Time analysis}

We query the data structures $D^{\ell/2^i}(.)$ for $i=0,\ldots$ until
$\ell/2^i = \lg w-1$ (we may stop the recursion before reaching this point).
For each recursive step we spend constant time querying the dictionary. Thus 
the global query time is upper bounded by $O(\lg\frac{\ell}{\lg w})$.

\paragraph*{Space analysis}

The space can be proved to be $O(n\lg u)$ bits by induction. Let us first
focus on the storage of the components $(m,r_m,M,r_M)$ of the dictionaries,
which need $\ell$ bits each.
For the base case $\ell = \lg w-1$ we have that $t$ keys are encoded using $O(t\lg w)$ bits. Now for any recursive data structure $D^\ell(S)$ we notice that the substructures $D^{\ell/2}(\hat S_p)$ are disjoint. Let us call $n_p = |\hat S_p|$ and $n=|S|$, then $\sum_p n_p \le n$.
We store the dictionary $I(P)$, which uses $O(n\ell)$ bits, and the substructures $D^{\ell/2}(\hat S_p)$.
 We denote by $s(\ell,|S'|)$ the space usage of any $D^\ell(|S|)$. Then the 
space usage of our $D^\ell(S)$ follows the recurrence 
$s(\ell,n)=\sum_p s(\ell/2,n_p)+O(n\ell)$. The solution to this recurrence is 
$O(n\ell) = O(n\lg u)$. 

In addition, the dictionaries store pointers $q$, whose size does not halve
from one level to the next. Yet, since each of the $n$ elements is stored in 
only one structure $D(\cdot)$, there are at most $n$ such structures and 
pointers to them. As the rest of the data occupies $O(n\ell)$ bits, we need
$n$ pointers of size $O(\lg n + \lg \ell) = O(\lg u)$ bits.\footnote{In the
tuples we must avoid using $\lg u$ bits for null pointers. Rather, we use just
a bitmap (with one bit per tuple) to tell whether the pointer is null or not,
and store the non-null pointers in a separate memory area indexed by $rank$
over this bitmap.} Thus the space is $O(n\lg u)$ bits.
% DIDNT' SEE THIS \footnote{Plus $O(n \lg w)$, which
%is dominated by $O(n \lg u)$ unless $u < w$, in which case we can simply use
%the technique of Section~\ref{sec:bitparpred} simulating a shorter machine
%word.}

\subsection{Predecessor Queries on Short Keys}
\label{sec:bitparpred}

We now describe the base case of the recursion for $O(\lg w)$-bit keys.
Suppose that we have a set $S$ of $t$ keys, each of length $\ell=(\lg w)/2-1$.
Clearly $t\leq \sqrt{w}/2$. What we want is to do predecessor search for any
$x$ over the set $S$. For that we first sort the keys (in ascending order)
obtaining an array $A[1,t]$. Then we pack them in a block $B$ of $t(\ell+1)$
consecutive bits (this uses $t(\lg w)/2\leq \sqrt{w}(\lg w)/4\leq w$ bits,
which is less than one word) where each key is separated from the other by a
zero bit. That is, we store the element $A[i]$ in the bits $B[(i-1)(\ell+1)+1,i(\ell+1)-1]$ and store a zero at bit $B[i(\ell+1)]$.

We now show how to do a predecessor query for a key $x$ on $S$ in constant time. This is done in the following steps:

\begin{enumerate}
\item We first duplicate the key $x$, $t$ times, and set the separator bits. 
That is, we compute $X=(x\cdot (0^\ell 1)^t) ~\textsc{or}~ (10^\ell)^t$. 
\item We subtract $B$ from $X$, obtaining $Y=X-B$.
This does in parallel the computation of $x-A[i]$ for all $1\leq i\leq t$, and 
the result of each subtraction (negative or nonnegative) is stored in the 
separator bit $Y[i(\ell+1)]$.
\item We mask all but separator bits. That is, we compute 
$Z=Y~\textsc{and}~(10^\ell)^t$.
\item We finally determine the rank of $x$. If $Z=0$ then we answer $(0,0)$. 
Otherwise, to find the first 1 in $Z$, we create a small universal mmphf 
storing the values $\{ 2^{\ell i},~ 1 \le i \le t\}$, which takes constant
time and $O(t \lg w) = O(\sqrt{w}\lg w) = o(w)$ bits. With the position of
the bit we easily compute the rank $r$ and extract the answer $y$ from the
corresponding field in $B$, so as to answer $(y,r)$.
\end{enumerate}

\subsection{Reducing Space Usage}

We now describe how the space usage can be improved to $O(n\lg(u/n))$. For 
this we use a standard idea. We partition the set $S$ into 
$n'=2^{\lfloor \lg n \rfloor}$ partitions using the $\lg n'$  most significant bits. For all the keys in a partition $S_p$, we have that the $\lg n'$ most significant bits are equal to $p$. 
Let $\hat S_p$ denote the set that contains the elements of $S_p$ truncated to their $\lg u-\lg n'$ least significant bits. We now build an independent predecessor data structure $D^{\lg u-\lg n'}(\hat S_p)$. Each such data structure occupies at most $c(|S_p|(\lg u-\lg n'))$ bits, for some constant $c$. We compact all those data structures in a memory area $A$ of $cn$ cells of $\lg u-\lg n'$ bits.
 
A bitvector $B[1,n+n']$ stores the size of the predecessor data structures. That is, 
for each $p\in[1,n']$ we append to $B$ as many 1s as the number of elements 
inside $S_p$, followed by a $0$. Then, to compute the predecessor of a key $x$ in $S$, we first compute $p=h(x)$ (here $h(x)$ extracts the $\lg n'$ most
significant bits and $l(x)$ the $\lg u - \lg n'$ least significant bits). 
Then we compute $r_0=select_0(B,p)-p$, which is
the number of elements in $S_q$ for all $q<p$. Then we query $D^{\lg u-\lg n'}(\hat S_p)$ (whose data structure starts at
$A[c\cdot r_0(\lg u-\lg n')]$) for the key $l(x)$, which returns an answer 
$(y,r)$. We now have two cases:
\begin{enumerate}
\item If the returned answer is $(y,r)\not=(0,0)$, then
the final answer is just $(pn'+y,r_0+r)$.
\item Otherwise, the rank of the answer is precisely $r_0$, but we must find 
the set $S_{p'}$ that contains it in order to find its value. 
There are two subcases:
\begin{enumerate}
	\item If $r_0=0$, then there is no previous element and we return $(0,0)$.
	\item Else we compute the desired index, $p'=select_1(r_0)-r_0$, and
query $D^{\lg u-\lg n'}(\hat S_{p'})$ for the maximum possible key, 
$1^{\lg u-\lg n'}$. This must return a pair $(y,r)$,
and the final answer is $(p'n'+y,r_0)$.
\end{enumerate}
\end{enumerate}

Since $B$ has $O(n)$ bits, it
is easy to see that the data structure occupies $O(n(\lg u-\lg n))$ bits and
it answers queries in time $O(\lg\frac{\lg u-\lg n}{\lg w})$.
We thus have proved the following theorem:

\begin{theorem} \label{thm:smallpred}
Given a set $S$ of $n$ keys over universe $[1,u]$, there is a data structure that occupies $O(n(\lg(u/n)))$ bits of space and answers predecessor queries 
in time $O(\lg\frac{\lg(u/n)}{\lg w})$.
\end{theorem}

\end{document}